\documentclass[aps,pre,twocolumn,superscriptaddress]{revtex4-2}
\usepackage{amssymb}
\usepackage{amsmath,bm}
\usepackage{graphicx,color}
\usepackage{bbm}
\usepackage[bottom]{footmisc}
\usepackage{multirow}
\usepackage{xcolor}
\usepackage{xpatch} 
\makeatletter   
\usepackage{xpatch} 

\usepackage{todonotes}

\newcommand{\B}[1]{{\bm{#1}}}

\usepackage{hyperref}
\newcommand{\figref}[1]{Fig.~\ref{#1}}

\newcommand{\rout}{r_\text{out}}

\newcommand{\appref}[1]{App.~\ref{#1}}

\begin{document}
\title{Continuum mechanics of differential growth in disordered granular matter}

\author{Noemie S. Livne}
\affiliation{Racah Institute of Physics, The Hebrew University of Jerusalem, Jerusalem, Israel 9190}
\author{Tuhin Samanta}
\affiliation{Dept. of Chemical Physics, The Weizmann Institute of Science, Rehovot 76100, Israel}
\author{Amit Schiller}
\affiliation{Racah Institute of Physics, The Hebrew University of Jerusalem, Jerusalem, Israel 9190}
\author{Itamar Procaccia} 
\affiliation{Dept. of Chemical Physics, The Weizmann Institute of Science, Rehovot 76100, Israel}
\affiliation{Sino-Europe Complexity Science Center, School of Mathematics, North University of China, Shanxi, Taiyuan 030051, China.}
\author{Michael Moshe}
\email[Corresponding author: ]{michael.moshe@mail.huji.ac.il}
\affiliation{Racah Institute of Physics, The Hebrew University of Jerusalem, Jerusalem, Israel 9190}

\date{\today}
\begin{abstract}
Disordered granular matter exhibits mechanical responses that occupy the boundary between fluids and solids, lacking a complete description within a continuum theoretical framework. Recent studies have shown that, in the quasi-static limit, the mechanical response of disordered solids to external perturbations is anomalous and can be accurately predicted by the theory of “odd dipole screening.”
In this work, we investigate \emph{responsive} granular matter, where grains change size in response to stimuli such as humidity, temperature, or other factors. We develop a geometric theory of odd dipole-screening, incorporating the growth field into the equilibrium equation. Our theory predicts an anomalous displacement field in response to non-uniform growth fields, confirmed by molecular dynamics simulations of granular matter. Although the screening parameters in our theory are phenomenological and not derived from microscopic physics, we identify a surprising relationship between the odd parameter and Poisson’s ratio. This theory has implications for various experimental protocols, including non-uniform heating or wetting, which lead to spatially varying expansion field.
\end{abstract}

\maketitle 
A densely packed assembly of elastic grains, forming a jammed granular material within a confined space, is able to withstand external shear stress akin to solid materials. Unlike conventional solids, stress in this state is transmitted through force chains, where close to the jamming point only a fraction of the particles actively bear the external load \cite{96JNB}. Consequently, when grains undergo volume expansion due to thermal, humidity \cite{20OH}, or other environmental changes, the mechanical impact varies depending on its position within the force chain network \cite{liu1994sound}. Therefore, the mechanical state of a responsive and growing granular matter is expected to significantly deviate from that of elastic-like growing solids \cite{goriely2005differential,goriely2008elastic,el2005self}. The main objective of this work is to study the effect of growth on granular matter within a continuum framework. 

A central objective in the theory of granular matter is the development of a complete continuum framework that accurately describe the corresponding complex  phenomenology. Attempts along this line include the description of active and passive granular matter in, or at the verge, of mechanical equilibrium \cite{blumenfeld2004stresses, nampoothiri2022tensor, huang2023odd, bera2024soft, vaibhav2024experimental,blumenfeld2024granular}. 
Despite the intricate microscopic details, recent studies have demonstrated that the \emph{equilibrium response} of jammed disordered solids to non-uniform strains can be effectively described at the continuum level using the theory of mechanical dipole-screening \cite{SchreiberKeren2021, lemaitre2021anomalous, mondal2022experimental, mondal2023dipole}. This theory considers the adjustment of the reference state through a distribution of quadrupole elastic charges \cite{livne2023geometric}. When the nucleation energy of quadrupoles is high, such as under significant pressure, the theory aligns with the elastic-like behavior of disordered solids. Conversely, when the nucleation energy is low, screening is dominated by dipole elastic charges which are formed by a non-uniform distribution of quadrupoles. These extend the concept of dislocations to disordered solids \cite{livne2023geometric, 15MSK, moshe2015geometry, kupferman2015metric}.

A more recent advancement in the theory revealed that screening in the presence of a glassy energy landscape, typical of disordered granular matter, violates energy conservation. The extended theory, termed odd-dipole screening, has been experimentally confirmed in sheared granular materials \cite{cohen2023odd}. 
Therefore, to study the effect of growth on disordered granular matter, below we focus on incorporating growth fields into the framework of odd dipole-screening.

We find it instructive to draw a connection between growth in 2d granular matter and in thin elastic sheets. A thin elastic sheet that experiences differential growth, that is each material element grows isotropically with an area change $\Delta S = \phi(\mathbf{x}) \, \Delta S_0$, either develops residual stresses when forced to remain planar, or escapes into the third dimension to exchange stretching energy with bending energy \cite{efrati2013metric}. The Gaussian curvature of the emergent three-dimensional configuration is fully determined by the growth field, and is the source for residual stresses in the planar state. For example, a flat elastic disk of radius $r_{\rm out}$ that experiences a growth field $\phi_\mathrm{sphere}(\mathbf{x}) = \tfrac{4 c^2}{((r/R)^2 + c^2)^2}$ with a real parameter $c$, develops residual stresses characterized by a nontrivial displacement field. Upon release into 3D, stresses are washed out by adopting a geometry of a spherical cap with Gaussian curvature $1/R^2$ \cite{moshe2014plane}. 

In this Letter we study theoretically and numerically the effect of nonuniform growth fields on disordered granular matter. 
On the theoretical side, we extend the theory of odd dipole-screening to include the effect of growth fields, and we derive explicit predictions for the displacement fields in response to non-uniform growth. We perform multiple numeric simulations varying the growth field with the same physical conditions, or varying physical conditions with the same growth field. We compare our predictions with observations from molecular dynamics simulations of disordered granular matter that experience differential growth, and find excellent agreement between theory and simulations. Importantly, we observe a characteristic property of odd screening, that is the breakdown of chiral symmetry with a non-vanishing tangential component of the displacement field. 
A central component of the theory is the emergence of two screening moduli, accompanying the classical elastic moduli. While we do not derive either elastic or screening moduli from microscopic interactions, we uncover an interesting and non-trivial relation between the observed ones. 
   
\begin{figure}
	\centering
	\includegraphics[width=\linewidth]{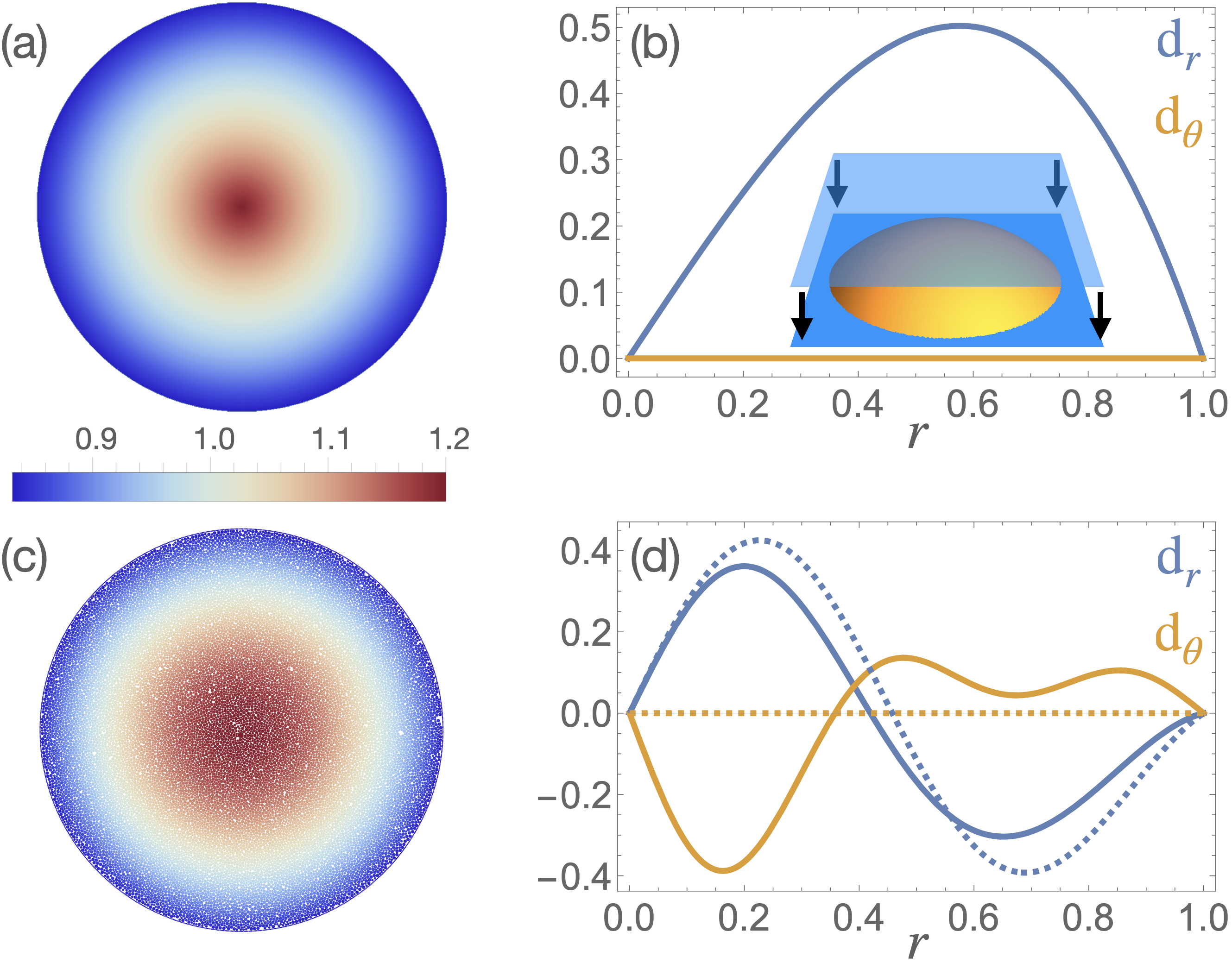}
	\caption{Growth-induced residual stresses: (a) A non-uniform growth profile $\phi_\mathrm{sphere}(\mathbf{x})$ inducing a mechanical state equivalent to flattening a spherical cap. (b) The resultant displacement field in the elastic case. (c) Illustration of the same growth field in a solid-like disordered granular structure. (d) Possible displacement fields corresponding to dipole-screening (dashed) and odd-dipole-screening (solid).}
	\label{kvetch}
\end{figure}

In this work we use the differential growth protocol that induces a reference state corresponding to a spherical geometry. The process is illustrated in Fig.~\ref{kvetch}, with the corresponding displacement fields in the elastic, screened, and odd-screened cases. 
To analyze the mechanical response in this setup, we start by reviewing the theory of odd dipole-screening and extending it to  account for differential growth.

In a series of recent papers, it was shown that the response to external loads in amorphous solids, result in the creation of a quadrupolar field $Q^{\alpha\beta}$, which modifies the rest state of the system to release strains \cite{livne2023geometric}. When the quadrupolar field is non-uniform, its divergence induces effective dipoles \cite{kupferman2015metric,moshe2015geometry,moshe2014plane}

\begin{equation}
	\B P_\mathrm{dip} \equiv \mathrm{div} \B Q \;.
	\label{defP}
\end{equation}
When the nucleation cost associated with the quadrupolar field is negligible, these dipoles dominate the mechanical screening mechanism. In this case the displacement field satisfies an equilibrium equation of the form 
\begin{eqnarray}
\B	\Delta \mathbf{d} &+& \frac{1+\nu}{1-\nu} \B \nabla (\B \nabla \cdot \mathbf{d}) + \B P_\mathrm{dip} = 0\notag \\
\B P_\mathrm{dip} &=&  \B \Gamma \,\mathbf{d} \;.
	\label{Equilibrium}
\end{eqnarray}
The second equation is obtained by integrating the equilibrium equation for $Q$, and set a gauge for $\mathbf{d}$ \cite{livne2023geometric}. The parameter $\nu$ is the 2D Poisson's ratio and $\Gamma$ the screening tensor, relating the inducing displacement with an induced dipole charge. 
Homogeneity and isotropy implie $\B \Gamma=\kappa\, \B I$, that is the induced dipole and the inducing displacement are co-linear.   Conventional elasticity is recovered in the limit $\kappa \to 0$, corresponding to a large dipole nucleation energy. 
 
In recent work \cite{cohen2023odd} it was shown that the rough energy landscape that characterizes disordered solids with mutliple local equilibria, adds an anti-symmetric term to $ \B \Gamma$, thus extending the theory to \emph{odd-dipole screening}. While still consistent with homogeneity and isotropy, it was shown that odd dipole screening is accompanied by spontaneous breakdown of chiral symmetry, and  accounts for the amount of work that is extracted or loaded into the material upon completing a closed trajectory in configuration space.
The screening tensor $\B \Gamma$ in that case is expressed as 
\begin{equation}
	\Gamma= \Big(\begin{matrix}
		\kappa_{\rm e} & -\kappa_{\rm o}\\
		\kappa_{\rm o} & \kappa_{\rm e} 
	\end{matrix}\Big)\equiv \kappa \Big(\begin{matrix}
	\cos\theta_k & -\sin\theta_k\\
	\sin\theta_k & \cos\theta_k
\end{matrix}\Big) \ .
\label{odd}
\end{equation}
Here $\kappa \equiv \sqrt{\kappa^2_{\rm e}+\kappa^2_{\rm o}}$, and $\tan \theta_k=\kappa_{\rm o}/\kappa_{\rm e}$. The screening parameter $\kappa>0$ and the odd-phase $\theta_{k}$ quantify the screening strength and the angle between the inducing displacement and the induced dipoles.
It is interesting to note that similar anti-symmetric terms exists in the electrostatic analog of non-hermitian dielectrics. In that case the dielectric tensor consists of an anti-symmetric term that quantifies energy loss or gain \cite{landau2013electrodynamics, friedland1980geometric}.
We emphasize that the odd dipole-screening mechanism developed in \cite{cohen2023odd} differs fundamentally from the recently studied odd-mechanics in driven granular matter \cite{huang2023odd}. While the former describes an a-symmetric constitutive relation between the displacement and dipoles, thus effectively violates translational symmetry, the latter preserves this translational symmetry and relates stresses with strains a-symmetrically. 
In what follows we generalize odd-dipole screening to account for differential growth.

The equilibrium equation \eqref{Equilibrium} reflects a force balance, and our task is to incorporate the effect of differential growth into this equation. The standard approach for deriving the governing equations in the presence of odd-mechanics is starting from an energy functional, deriving the equations of motion, and then incorporate odd-coupling terms that are consistent with the symmetries of the system. 
We repeat this methodology and start by deriving the equilibrium equation of a growing sample in the presence of even dipole screening. 

A key component of our derivation is the screened strain. While the full strain-tensor derived from the displacement field is $u = \tfrac{1}{2} \left(\nabla \mathrm{d}^T + \nabla \mathrm{d}\right)$, the screened strain for which elastic energy penalizes is modified by $q$, the anelastic strain, $u_\mathrm{el} = u -q$. This anelastic strain induces a quadrupolar elastic charge distribution $Q$ \cite{livne2023geometric}. Then, the energy penalizes not only for elastic strains, but also for the nucleation of quadrupoles and their gradients. When screening is dominated by dipoles, one recovers \eqref{Equilibrium}. When differential growth takes place, the reference state changes, and the screened strain is $u_\mathrm{el} = u - (\phi(\mathrm{x})-1) \B{I} - q$. 
The equilibrium equation is (see \appref{app:der})
\begin{equation}
	\Delta \mathbf{d} + \frac{1+\nu}{1-\nu} \nabla \left( \nabla \cdot \mathbf{d} -\phi(\mathbf{x}) \right) + \Gamma \mathbf{d} = 0.
	\label{eq:Equilibrium_with_expansion}
\end{equation}
Like before, the effect of odd coupling is incorporated at the level of equation of motion with $\B\Gamma$ containing an anti-symmetric term.

Next, we consider an expansion field in the form
\begin{equation}
	\phi  = \frac{4 c^{2}}{\left(c^{2}+(r / R)^{2}\right)^{2}} \ , \quad c=\sqrt{4-\left(\frac{r_{\text {out }}}{R}\right)^{2}} \ .
	\label{inflation}
\end{equation}
In the limit $r_{\text{out}}\ll R$, $\phi \to 1$ and the equation reduces back to its original form \eqref{Equilibrium}. For finite $R$, the choice of $c$ as written above guarantees that the total reference area (and therefore the pressure) remains invariant during growth. If this expansion field were applied to an elastic sheet, it would induce a surface with constant curvature $\frac{1}{R^{2}}$. However, the sheet would develop residual stresses when constrained to remain in a plane \cite{moshe2014plane}. To predict the stress and displacement fields in the presence of odd-dipole screening, we use the polar symmetry of the problem to express the displacement field as $\mathbf{d}=\left(d_{r}(r), d_{\theta}(r)\right)$. In the limit of $r_{\text{out}}\ll R$, the solution to \eqref{eq:Equilibrium_with_expansion} takes the form
\begin{eqnarray}
	d_r(r) &=& \alpha_r  J_1(r/l_1) + \beta_r J_1(r/l_2) + \gamma_r r \;, \notag\\
	d_\theta(r) &=& \alpha_\theta  J_1(r/l_1) + \beta_\theta J_1(r/l_2) + \gamma_\theta r \;,
\end{eqnarray}
with the parameters $\alpha_i, \beta_i, \gamma_i$ and $l_i$ functions of the screening parameters, Poisson's ratio, imposed radius of curvature $R$, and system size $\rout$ (for explicit expressions see App.~\ref{app:Sol}).
Sketches of representative solutions for even ($\theta_\kappa = 0$) and odd ($\theta_\kappa \neq 0$) screening modes are depicted in \figref{kvetch}(d) in dashed and solid lines respectively. 

\begin{figure}
	\centering
	\includegraphics[width=\linewidth]{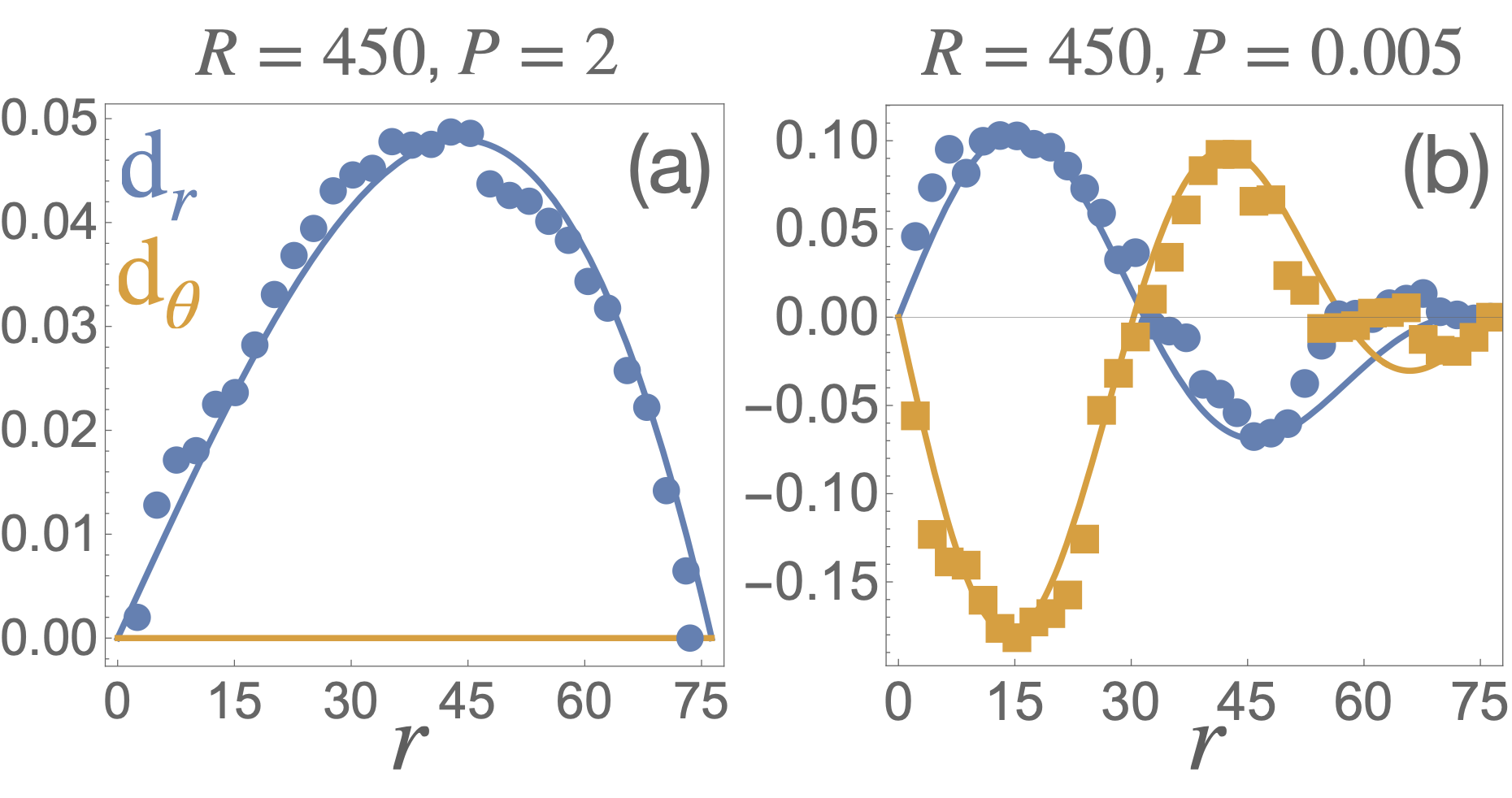}
	\caption{The angle-averaged displacement fields induced by the nonuniform growth field $\phi_\mathrm{sphere}$ corresponding to radius of curvature $R_\mathrm{curv} = 450$. (a) Quasi elastic response at high pressure $P = 2$. (b) Odd-dipole-screening at a lower pressure $P = 0.005$. Simulation data is represented by discrete markers, and the theoretical prediction fit by solid lines.}
	\label{fig:RepDisp}
\end{figure}

To test our theoretical predictions we performed numerical simulations of disordered granular matter consisting of frictionless disks. In our simulations we started from an unjammed state and reduced the outer radius of the domain to achieve a target  pressure. In each simulation we imposed a growth field and allowed the system to relax until it reached a new equilibrium state. 
The displacement field is measured between the equilibrium state prior to growth to that after it. 
To compare the measured displacement field with our predictions we perform an angle-averaging of the radial and tangential displacement components of each particle. 

\begin{figure}
	\centering
	\includegraphics[width=\linewidth]{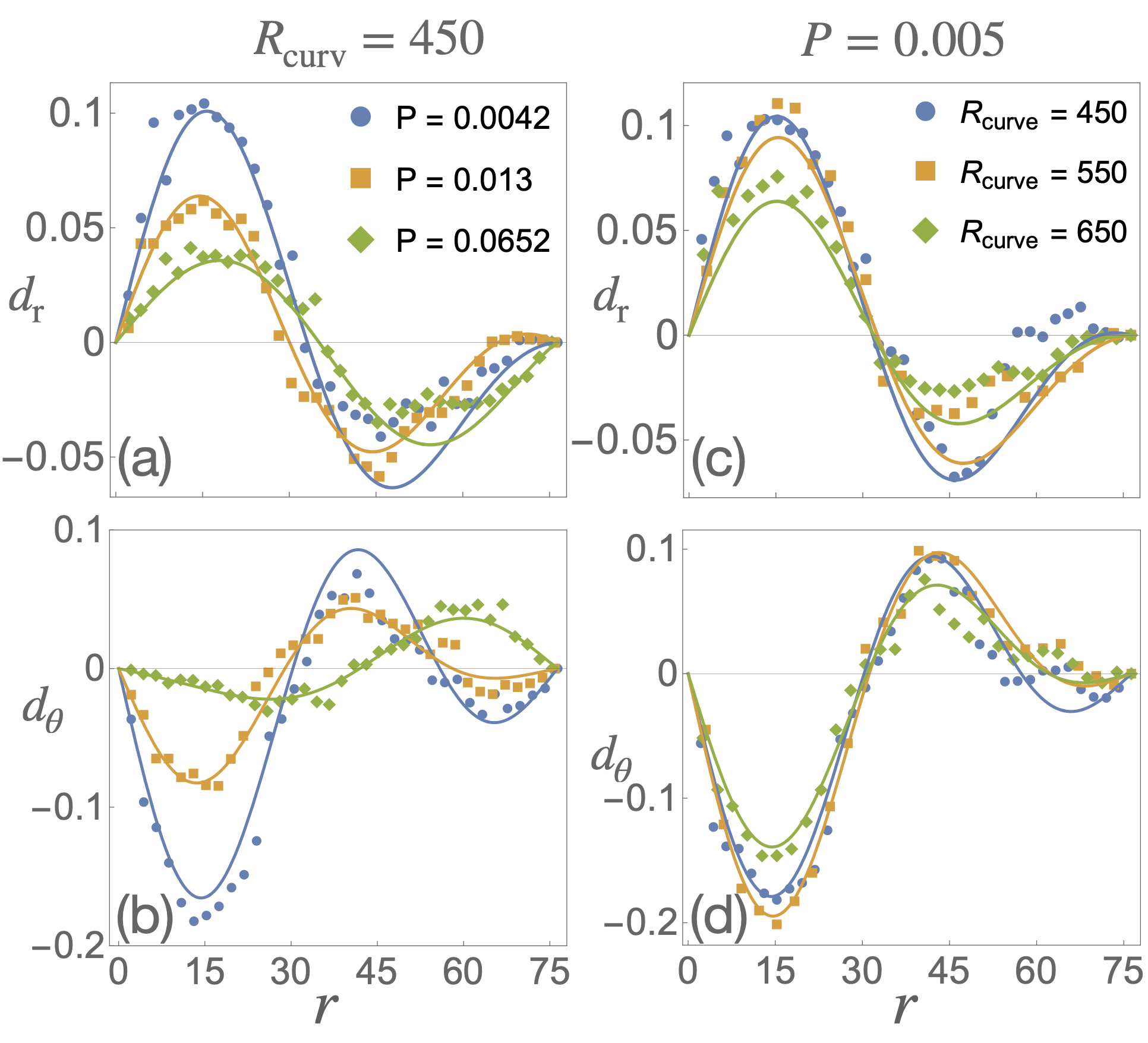}
	\caption{Left panel: Comparison between simulations and theory of $d_r$ (a) and $d_\theta$ (b) for a range of pressures with fixed imposed radius of curvature $R_\mathrm{curv}$. Right panel: Comparison between simulations and theory of $d_r$ (c) and $d_\theta$ (d) for a range of $R_\mathrm{curv}$ with fixed pressure.}
	\label{fig:CurvatureControlled}
\end{figure}

\begin{figure}
	\centering
	\includegraphics[width=\linewidth]{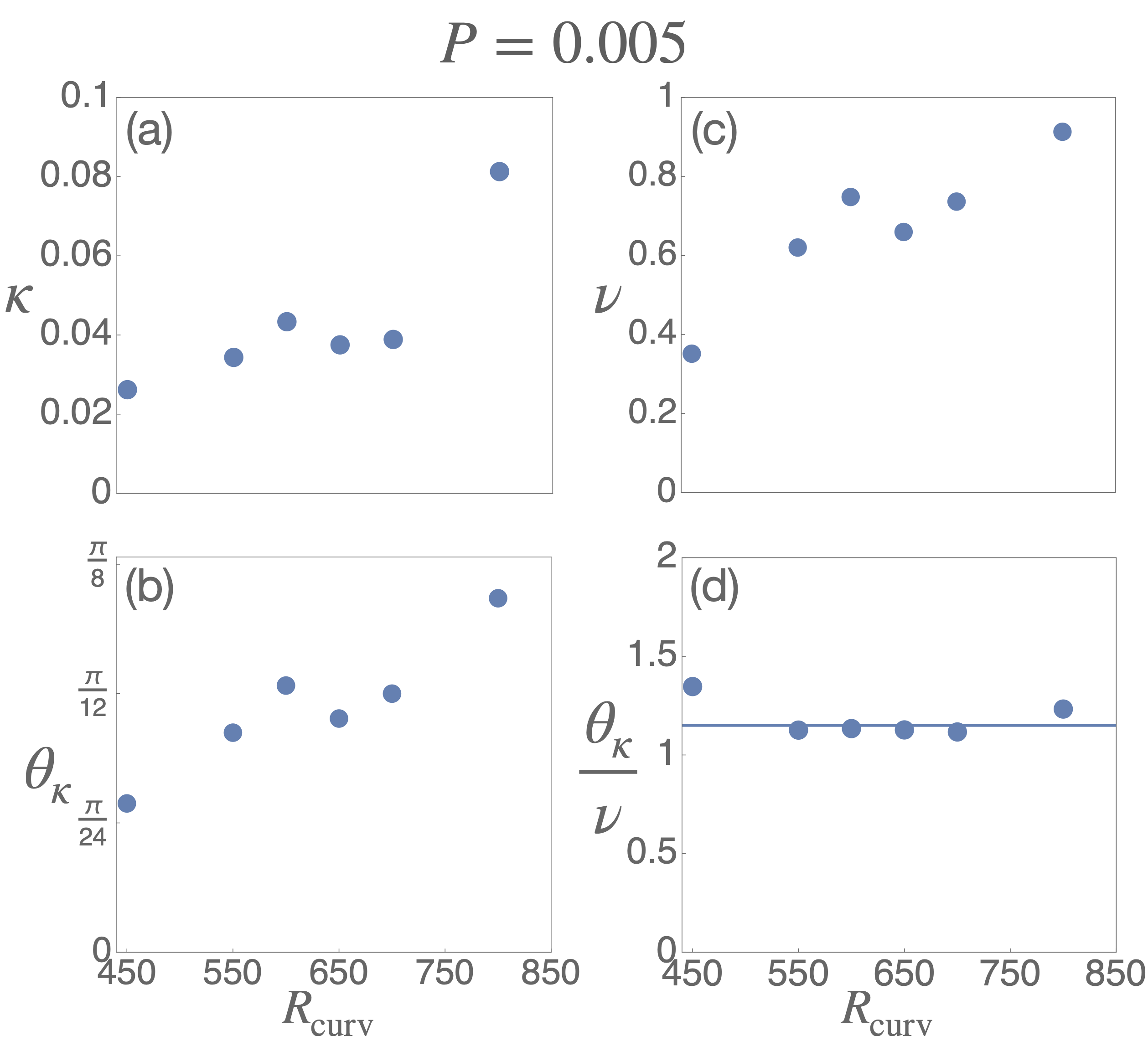}
	\caption{The fitting parameters as function of imposed curvatures $R_\mathrm{curv}$, with fixed pressure $P$. Shown are (a) the screening parameter $\kappa$, (b) the odd-phase $\theta_\kappa$, (c) Poisson ratio $\nu$, and (d) the odd-phase divided by the Poisson ratio. Note in (d) the appearance of a near-constant ratio around $1.1$ (denoted by a line).}
	\label{fig:Fit-params_CurvatureControlled}
\end{figure}

\begin{figure}
	\centering
	\includegraphics[width=\linewidth]{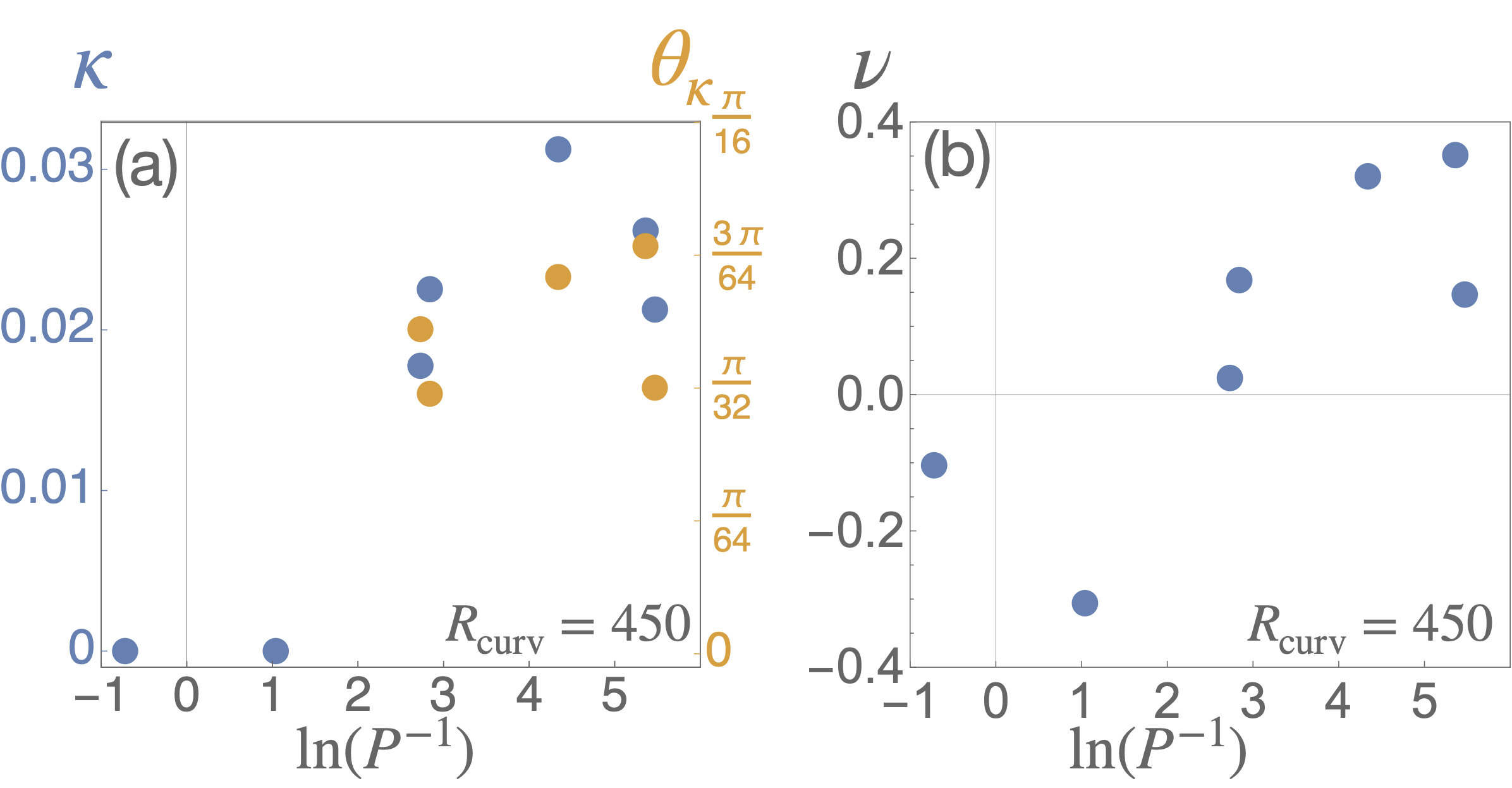}
	\caption{The fitting parameters as function of pressure $P$, with  fixed curvature $R_\mathrm{curv}$. Shown in (a) are the screening parameter $\kappa$ in blue on the left axis, and the odd-phase $\theta_\kappa$ in orange on the right axis. (b) Poisson's ratio, indicating a sign change at the onset of screening.}
	\label{fig:Fit-params_AreaFracControlled}
\end{figure}

Dipole screening in general, and odd-screening in particular, are expected to take place when quadrupoles nucleation costs are very low. We therefore expect elastic-like behavior to take place, for example, at high pressures and odd-dipole screening at low pressures.  
In Fig.~\ref{fig:RepDisp} we show two typical examples of displacement fields at low and high pressures. The comparison between theory and simulations require a simultaneous fit for $d_r$ and $d_\theta$ with respect to three fitting parameters: $\kappa, \theta_\kappa$ and $\nu$. We see that theory and simulations are in excellent agreement with our expectation and with the theoretical predictions.  

In the high pressure case the response is quasi-elastic as shown in Fig.~\ref{fig:RepDisp}(a). In this case the fitting parameters are $\kappa = 0$ and $\nu = -0.1$. 
In the low pressure case the response is anomalous, and it breaks chiral symmetry with a non-vanishing tangential displacement,  as shown in Fig.~\ref{fig:RepDisp}(b). 
The fitting parameters are $\kappa = 0.026$, $\theta_\kappa \approx \pi/6 $ and $\nu = 0.35$. We see that the effect of odd coupling is moderate, with an angle of $\pi/6$ between the inducing displacement and induced dipole. 

We continue with a systematic study of the anomalous response to non-uniform growth, and its dependence on the controlled parameters: the radius of curvature $R_\mathrm{curv}$ and the pressure $P$. 
We start by varying the pressure in systems with a fixed growth field that correspond to radius of curvature $R_\mathrm{curv}=450$. Three representative examples are shown in left panel of  Fig.~\ref{fig:CurvatureControlled}(a,b), showing that pressure controls the level of screening. For example, we see that the tangential displacement component decreases in amplitude as the pressure increases. 
We continue by varying the radius of curvature $R_\mathrm{curv}$ in systems with fixed pressure $P = 0.005$. Three representative examples are shown in right panel of Fig.~\ref{fig:CurvatureControlled} (c,d), showing that curvature also controls the level of screening. For example, we see that the  displacement amplitude decreases as the radius of curvature increases, as expected. 
In both curvature-controlled or pressure-controlled systems, the agreement between theoretical predictions and observations is very good. Interestingly, it seems that the odd coupling is also affected by the pressure and radius of curvature. To better quantify this impression we plot the fitted moduli as function of the controlled parameter. For example, in the case of pressure controlled simulations, we plot $\kappa$, $\theta_\kappa$ and $\nu$ as function of the imposed radius of curvature $R_\mathrm{curv}$, as shown in \figref{fig:Fit-params_CurvatureControlled}. 

In this figure we see explicit dependence of the screening moduli and Poisson's ratio on the imposed radius of curvature. In principle, the screening parameters and the Poisson's ratio are emergent properties that depend on microscopic properties, e.g. particle size ratios. The dependence of these emergent properties can be a complicated function of the microscopic parameters. While currently we do not have a derivation of such functions, we observe that the ratio between the odd phase and the Poisson's ratio is nearly constant, $\theta_\kappa/\nu = 1.1$. 
Investigation of these relations, and understanding which microscopic parameters determine them, requires further numeric simulations and experiments with varying microscopic properties, such as particle size ratios and more. Such an investigation is left for future research.

Next we study the dependence of screening moduli on the imposed pressure for fixed $R_\mathrm{curv}$. 
For presentation purposes we plot the screening moduli as function of $\Phi = \ln(P^{-1})$, see \figref{fig:Fit-params_AreaFracControlled}.
We find that at large pressures (low $\Phi$) the effect of dipole screening disappear, with quasi-elastic behavior and $\kappa=0$. In this regime $\theta_\kappa$ is meaningless. 
At lower pressures screening mechanism is dominated by odd-dipoles, and finite values for $\kappa$ and $\theta_\kappa$ are observed. This implies the possibility of a transition or a crossover from a quasi-elastic regime to an odd-dipole screening regime \cite{jin2024intermediate,zaccone2023theory}.

In summary, in this work we have shown that the recently developed theory of odd-dipole-screening \cite{cohen2023odd} accurately predicts the functional form of the displacement field in responsive granular matter. A main feature of the theory is that it describes materials that do not conserve energy. The success of the theory implies that the numerical model of the disordered granular assembly do not conserve energy, as expected from a material with glassy energy landscape. 

As a linear theory, the theory is expected to hold for small deformations. Therefore, we limited the numerical simulations to small growth fields $\phi_\mathrm{sphere}$, that is, the radius of curvature induced by the growth is much larger than $r_{\rm out}$. We expect that when these become more comparable, nonlinear terms will need to be included into Eq.~(\ref{odd}), but this beyond the scope of the present Letter. 

The growth protocol studied in this letter is a continuum generalization of the particle-inflation protocol studied in previous works on anomalous mechanics in disordered solids (e.g. \cite{mondal2022experimental, fu2024odd}). There, the quasi-elastic response to a single-particle inflation had the property that it depended only on geometric properties. Therefore, elastic moduli could not be extracted in the quasi-elastic regime. Here we showed that the quasi-elastic response to the growth protocol $\phi_\mathrm{sphere}$ depends on the Poisson's ratio. The numerical simulations revealed that the Poisson's ratio is negative in the quasi-elastic regime, and positive in the odd-screening regime. This suggests that the Poisson's ratio may serve as an indicator for the onset of dipole screening.

Last but no the least, we note that the growth protocol in this work was fully prescribed. In reality, growth and mechanics are strongly coupled \cite{gniewek2019biomechanical}. Our theory put the foundations for a future theory that couples the dynamics of the growth field and the mechanical state of the growing system.


{\bf Acknowledgments:}  NL TS and AS contributed equally to this work: NL developed the theoretical framework by combining non-euclidean elasticity with the theory of odd-dipole screening. AS conceived the theoretical problem and developed the theory in its fist stages. TS performed numerical simulations. NL and TS analyzed the data. NL, TS, IP and MM wrote the manuscript. MM NL and AS have been supported by Israel Science Foundation Grant No. 1441/19.
TS and IP have been supported by the joint grant between the Israel Science Foundation and the National Science Foundation of China, and by the Minerva Foundation, Munich, Germany. 


\bibliography{refs}

\clearpage

\appendix

\section{The expansion profile}

Consider an elastic manifold shaped as a planar disc of radius $r_{\text{out}}$, 
which undergoes a local plastic deformation modeled as an expansion
field applied to each point in the disc. 

We want to pick a function $\phi\left(r\right)$ so that the metric $\bar{g}_{0}=\phi\left(r\right) \left(\begin{smallmatrix}1 & 0\\
	0 & r^{2}
\end{smallmatrix}\right)$ has constant positive Gaussian curvature, $K_{G}\left(\bar{g}_{0}\right)=\frac{1}{R^2}$.
We denote $\phi\left(r\right)=e^{2\varphi\left(r\right)}$, and note that the Gaussian curvature corresponding to this form of $\bar{g}_{0}$ is $-e^{2\varphi\left(r\right)}\Delta \varphi\left(r\right)$,
so we have to solve
\begin{equation}
	-e^{2\varphi\left(r\right)}\Delta \varphi\left(r\right)=\frac{1}{R^2}.    
\end{equation}
The solution to this equation is
\begin{equation}
	e^{2\varphi\left(r\right)}=\frac{4c_{1}^{2}c_{2}^{2}r^{2c_{1}-2}}{\left(\frac{1}{R^2} r^{2c_{1}}+c_{2}^{2}\right)^{2}}
\end{equation}
where $c_{1},c_{2}$ are arbitrary constants. We are applying the
profile to a disc, so the value at the origin should not diverge.
This requires $c_{1}=1$, leaving us with
\begin{align}
	\phi\left(r\right) & =e^{2\varphi\left(r\right)}=\frac{4c_{2}^{2}}{\left(\left(\frac{r}{R}\right)^2 +c_{2}^{2}\right)^{2}}.
\end{align}
If an elastic body was given this metric and allowed to reach an equilibrium
configuration in three-dimensional space, it would form a spherial
cap. The area of this cap would be 
\begin{equation}
	\intop_{0}^{r_{\text{out}}}\intop_{0}^{2\pi}\sqrt{\det\left(\bar{g}_{0}\right)}d\theta dr=\frac{4\pi r_{\text{out}}^{2}}{c_{2}^{2}+\left(\frac{r_{\text{out}}}{R}\right)^2 }.
\end{equation}
In order to keep
the pressure on the system constant, we choose $c_{2}$ such that
the deformation doesn't change the total area:
\begin{equation}
	\frac{4\pi r_{\text{out}}^{2}}{c_{2}^{2}+\left(\frac{r_{\text{out}}}{R}\right)^2}=\pi r_{\text{out}}^{2}\implies c_{2}=\sqrt{4-\left(\frac{r_{\text{out}}}{R}\right)^2 }.
\end{equation}
This leaves us with
\begin{equation}
	\phi\left(r\right)=\frac{4\left(4-\left(\frac{r_{\text{out}}}{R}\right)^2\right)}{\left(4+\left(\frac{r}{R}\right)^2-\left(\frac{r_{\text{out}}}{R}\right)^2\right)^{2}}.
\end{equation}

\section{Derivation of the equilibrium equation}
\label{app:der}
To derive the equilibrium equation we define an energy density that
takes into account the different types of response: the particle rearrangements
and the elastic deformation. The equations will be given in terms
of a displacement field, but in order to account for the expansion
profile and the local rearrangements we will use a geometric formalism.

In geometric terms, applying the expansion profile corresponds to
assigning a new reference metric $\bar{g}_{0}=\phi(\mathbf{x})\eta$,
where $\eta$ is the Euclidean metric that described the flat two-dimensional
sample before it was deformed. The sample then responds through particle
rearrangements and elastic deformation, reaching a new equilibrium
configuration. This configuration induces a new Euclidean metric $g$
on the body manifold of the sample. Since the initial and final configurations
are both flat, the original metric $\eta$ and the actual metric $g$
are related by a well-defined displacement field $\mathbf{d}$, through
the total strain field $u_{\text{tot}}=\frac{1}{2}(g-\eta)=\frac{1}{2}\left(\nabla\mathbf{d}+\nabla\mathbf{d}^{T}\right)$. 

The energetic cost of an elastic deformation is due to the elastic strain, which we will define as a geometric strain \cite{efrati2013metric}.
The deformed configuration is described by the metric $\bar{g}_{0}$, but also undergoes anelastic particle rearrangements. As in \cite{livne2023geometric},
we define a temporary reference metric $\bar{g}=\bar{g}_{0}+q$ where
$q$ the anelastic strain describing the particle
rearrangements, and is both traceless (area conserving) and
symmetric. The elastic strain is then the deviation of the sample's
actual metric $g$ from this effective reference metric $\bar{g}$:
\begin{equation}
	u_{\text{el}}=\frac{1}{2}\left(g-\bar{g}\right)=\frac{1}{2}\left(g-\bar{g}_{0}-q\right).\label{eq:u_el}
\end{equation}
Assuming small strains, we define a Hookean elastic energy density,
$U_{\text{el}}=\frac{1}{2}\mathcal{A}u_{\text{el}}^{2}$, where $\mathcal{A}$
is the elastic tensor encoding material properties. 

We further take into account the nucleation cost of the quadrupolar
perturbation field. As discussed in \cite{livne2023geometric}, if the pressure applied to the system is such that the nucleation cost of a single quadropolar charge is negligible, a non-uniform field of quadrupolar charges induces an effective field of dipoles $\mathbf{P}_\mathrm{dip}=\text{Div}Q$ with non-negligible energetic cost. Here the quadrupole charge field relates to the anelastic strain through $Q^{\alpha\beta} = \varepsilon^{\alpha\mu}\varepsilon^{\beta\nu} q_{\mu\nu}$.
The nucleation energy will have the form $U_{p}=\frac{1}{2}\Lambda_{P}P_\mathrm{dip}^{2}$,
where $\Lambda_{P}$ is a tensor that encodes material properties. Finally,
we add to the energy density the correction term $U_{q}=-\frac{Y}{8\left(1+\nu\right)}q^{2}$, 
otherwise the nucleation cost of the quadrupolar field won't be negligible compared
to $U_{p}$ (this is due to the appearance of $q$ in \eqref{eq:u_el}).
Finally, we have the total energy density 
\begin{align}
	\mathcal{U} & =U_{\text{el}}+U_{q}+U_{p}\\
	& =\frac{1}{2}\mathcal{A}u_{\text{el}}^{2}-\frac{Y}{8\left(1+\nu\right)}q^{2}+\frac{1}{2}\Lambda_{P}P_\mathrm{dip}^{2}.
\end{align}

For small strains, expansion profile, and plastic deformation fields
in a homogeneous and isotropic two dimensional material, the tensor
in the elastic term has the form 
\begin{equation}
	\mathcal{A}^{\alpha\beta\gamma\delta}=\frac{\nu Y}{1-\nu^{2}}\left(\eta^{\alpha\beta}\eta^{\gamma\delta}+\frac{1-\nu}{2\nu}\left(\eta^{\alpha\gamma}\eta^{\beta\delta}+\eta^{\alpha\delta}\eta^{\beta\gamma}\right)\right),
\end{equation}
where $Y$ is the Young's modulus, $\nu$ the Poisson ratio, and
$\eta$ the metric of the undeformed initial Euclidean system.
The tensor in the dipole screening term is 
\begin{equation}
	\Lambda_{P}=\frac{1}{\kappa_{p}}\eta
	\label{screening tensor symmetric}
\end{equation}
for some scalar $\kappa_{p}$. Finally, the energy density is integrated
over the material body manifold. In this limit the surface element
is the Euclidean surface element $dS_{\eta}$ and the energy functional
is
\begin{equation}
	F=\int\left(\frac{1}{2}\mathcal{A}u_{\text{el}}^{2}-\frac{Y}{8\left(1+\nu\right)}q^{2}+\frac{1}{2}\Lambda_{p}P_\mathrm{dip}^{2}\right)dS_{\eta}+B.T.
\end{equation}

The Euler-Lagrange equations of this energy functional, found by varying
$d$, are
\begin{align}
	\Delta\mathbf{d}+\frac{1+\nu}{1-\nu}\vec{\nabla}\left(\vec{\nabla}\cdot\mathbf{d}-\phi(\mathbf{x})\right)+\mathbf{P}_\mathrm{dip} & =0
	\label{eq:Equilibrium with expansion}
\end{align}
where $P_\mathrm{dip}^{\mu}=\nabla_{\nu}\varepsilon^{\alpha\mu}\varepsilon^{\beta\nu}q_{\alpha\beta}$, in which $\varepsilon$ is the anti-symmetric Levi-Civita tensor corresponding to the Euclidean metric $\eta$.
The equations obtained by varying $q=\left(\begin{smallmatrix}q_{11} & q_{12}\\
	q_{12} & -q_{11}
\end{smallmatrix}\right)$ are somewhat more cumbersome:
\begin{equation}
	\begin{aligned}
		\nabla_{1}d_{1}-\nabla_{2}d_{2}+\frac{2\kappa_{p}\left(1+\nu\right)}{Y}\Delta q_{11} & =0,\\
		\nabla_{2}d_{1}-\nabla_{1}d_{2}+\frac{2\kappa_{p}\left(1+\nu\right)}{Y}\Delta q_{12} & =0.
	\end{aligned}
	\label{eq:q_eqs}
\end{equation}

The displacement field is given to vanish at the boundary (the area of the sample isn't allowed to change). As for the polarization field, we are not interested at present in its boundary conditions, because we solve for $\mathbf{d}$ and can integrate out the $\mathbf{P}_\mathrm{dip}$ degrees of freedom: from \eqref{eq:q_eqs}, the $q$ degrees of freedom can be integrated out to find
\begin{equation}
	\mathbf{P}_\mathrm{dip}=\frac{2\kappa_{p}\left(1+\nu\right)}{Y}\left(\mathbf{d}-\mathbf{d}_{0}\right)
	\label{eq:P-d relation}
\end{equation}
where $\mathbf{d}_{0}$ are integration constants. Because they define
a solid translation, we can set it to zero. Plugging this into \eqref{eq:Equilibrium with expansion}, we have an equilibrium equation in terms of $\mathbf{d}$ alone
\begin{equation} 
	\Delta\mathbf{d}+\frac{1+\nu}{1-\nu}\vec{\nabla}\left(\vec{\nabla}\cdot\mathbf{d}-\phi(\mathbf{x})\right)+\frac{2\kappa_{p}\left(1+\nu\right)}{Y}\mathbf{d}=0.
\end{equation}
In the limit $\kappa_p \gg 1$ the dipole nucleation cost is negligible, as can be seen from \eqref{screening tensor symmetric}. In this limit, $\mathbf{P}_\mathrm{dip}$ fully screens the in-homogeneous contribution of $\phi(\mathbf{x})$, compensating for the curvature due to $\phi(\mathbf{x})$.

The generalization to odd-mechanical systems follows the same procedure as \cite{cohen2023odd}, since the only difference is the non-homogeneous contribution $\phi(\mathbf{x})$, and adds an anti-symmetric component to \eqref{eq:P-d relation}:
\begin{equation}
	\mathbf{P}_\mathrm{dip}=\left(
	\begin{matrix}\tilde{\kappa}_e & \tilde{\kappa}_o\\
		-\tilde{\kappa}_o & \tilde{\kappa}_e
	\end{matrix}
	\right) \mathbf{d},  
\end{equation}
where the tildes denote the constant factors that were incorporated into the constants. We can write this operation as a multiplication by a constant $\kappa$ and a rotation by angle $\theta_k$, giving us the equation
\begin{equation}
	\Delta \mathbf{d} + \frac{1+\nu}{1-\nu} \nabla \left( \nabla \cdot \mathbf{d} -\phi(\mathbf{x}) \right) + \kappa \big(\begin{smallmatrix}
		\cos\theta_k & -\sin\theta_k\\
		\sin\theta_k & \cos\theta_k
	\end{smallmatrix}\big) \mathbf{d} = 0,
	\label{eq:Equilibrium with expansion 2}
\end{equation}
which is equation \eqref{Equilibrium}.

\section{Analytical solution for the displacement field of ``flattened granular sphere''}
\label{app:Sol}
In the limit $r_{\text{out}}\ll R$, the leading order contribution of the non-homogeneous term due to $\phi(\mathbf{x})$ is proportional to $\frac{r^2}{R^2}$, and the equation is

\begin{equation}
	\Delta \mathbf{d} + \frac{1+\nu}{1-\nu} \nabla \left( \nabla \cdot \mathbf{d} +\frac{r^2}{2 R^2} \right) + \kappa \big(\begin{smallmatrix}
		\cos\theta_k & -\sin\theta_k\\
		\sin\theta_k & \cos\theta_k
	\end{smallmatrix}\big) \mathbf{d} = 0,
	\label{eq:Equilibrium with expansion at limit}
\end{equation}

The explicit solution that satisfies this equation with the stated boundary conditions is

\begin{widetext}

\begin{equation}
	\begin{aligned}
		d_{r}\left(r\right) & = \frac{C}{\kappa}\left\{ J_{1}\left(\omega_{m}r_{\text{out}}\right)
		\left[\kappa^{2}\left(\omega_{p}^{4}-\omega_{m}^{4}\right)
		J_{1}\left(\omega_{p}r_{\text{out}}\right)r
		+r_{\text{out}}\omega_{m}^{4}\left(\kappa^{2}-\omega_{p}^{4}\right)J_{1}\left(\omega_{p}r\right)\right]\right. \\
		& \left. \hphantom{=\frac{C}{\kappa} - } -r_{\text{out}}\omega_{p}^{4}\left(\kappa^{2}-\omega_{m}^{4}\right)
		J_{1}\left(\omega_{p}r_{\text{out}}\right)J_{1}\left(\omega_{m}r\right)\right\} \\
		d_{\theta}\left(r\right) & = C\sqrt{
			\left(\kappa^{2}-\omega_{m}^{4}\right)\left(\kappa^{2}-\omega_{p}^{4}\right)}
		\left\{ J_{1}\left(\omega_{m}r_{\text{out}}\right)\left[\left(\omega_{m}^{2}
		-\omega_{p}^{2}\right)J_{1}\left(\omega_{p}r_{\text{out}}\right)r
		-r_{\text{out}}\omega_{m}^{2}J_{1}\left(\omega_{p}r\right)\right] \right. \\
		&\hphantom{= C\sqrt{
				\left(\kappa^{2}-\omega_{m}^{4}\right)\left(\kappa^{2}-\omega_{p}^{4}\right)} - } \left.
		+r_{\text{out}}\omega_{p}^{2}J_{1}\left(\omega_{p}r_{\text{out}}\right)J_{1}\left(\omega_{m}r\right)\right\} \\
	\end{aligned}
\end{equation}
	
Where

\begin{equation}
	C = \frac{1}{R^{2}}
	\frac{\left(\kappa^{2}-\omega_{m}^{2}\omega_{p}^{2}\right)}
	{\kappa\omega_{m}^{2}\omega_{p}^{2}\left(\omega_{m}^{2}-\omega_{p}^{2}\right)
	\left(\kappa^{2}+\omega_{m}^{2}\omega_{p}^{2}\right)
	J_{1}\left(\omega_{m}r_{\text{out}}\right)J_{1}\left(\omega_{p}r_{\text{out}}\right)}
\end{equation}

and

\begin{equation}
	\begin{aligned}
		\omega_{m} & =\sqrt{\frac{\kappa}{2(\lambda+1)}} \sqrt{(\lambda+2) \cos 	\theta_{k}-\sqrt{(\lambda+2)^{2} \cos ^{2} \theta_{k}-4(\lambda+1)}} \\
		\omega_{p} & =\sqrt{\frac{\kappa}{2(\lambda+1)}} \sqrt{(\lambda+2) \cos 	\theta_{k}+\sqrt{(\lambda+2)^{2} \cos ^{2} \theta_{k}-4(\lambda+1)}}
	\end{aligned}
\end{equation}

\end{widetext}

\section{Simulation Protocol. }

For the flattening protocol, we employed four sizes of frictionless disks, of radii 0.4,0.5,0.6, and 0.7, placed randomly in a circular disk of initial radius $r_{out} =75$, all in dimensionless units. There are $N= 15000$ discs in all.  An initial area fraction below the jamming threshold was chosen, and then the outer radius was isotropically reduced to achieve a finite chosen pressure. Open source code Large-scale Atomic/Molecular Massively Parallel Simulator(LAMMPS) are used to perform the simulations. In these, the normal contact force is a Hertzian force with force-constant $K{_n} = 2000$ and the viscoelastic damping constant for normal contact $\gamma{_n} = 100$. The mass of each disk is $1$. The tangential contact force is zero since the system has no friction.
Using Newton's second law of motion with damping, the system is relaxed to mechanical equilibrium after each step, i.e., the total force on each disk is minimized to be smaller than $10^{-6}$. Once a mechanically stable configuration is reached at a desired pressure, each disc is inflated according to Eq.~(\ref{inflation}), and subsequently mechanical equilibration follows, and the resulting 
displacement field measured. The displacement field depends on both $r$ and $\theta$, and we
decompose it into the radial component $d_r( r,\theta)$ and the angular component $d_\theta( r,\theta)$. Finally, comparison with the theoretical functions as shown in Fig.~\ref{fig:RepDisp} is obtained by angle averaging, $d_r(r)\equiv \frac{1}{2\pi} \int d\theta~ d_r( r,\theta)$, and similarly for the tangential component.

\end{document}